%% file: session_sigir_202.tex
\documentclass[sigconf]{acmart}

\usepackage{xspace}
\usepackage{subfig}
\usepackage{enumerate}
\usepackage{enumitem}
\usepackage{color}
\usepackage{graphicx}
\usepackage{wasysym}
\usepackage{balance} 

\newtheorem{definition}{Definition}

\usepackage{threeparttable}
\usepackage{dcolumn}
\newcolumntype{d}[1]{D{.}{.}{#1}}

\newcommand{\eat}[1]{}
\newcommand{\paratitle}[1]{\vspace{1ex}\noindent \textbf{#1}}

\let\oldhat\hat
\renewcommand{\vec}[1]{\mathbf{#1}}
\renewcommand{\hat}[1]{\oldhat{\mathbf{#1}}}
\renewcommand{\matrix}[1]{\mathbf{#1}}

\newcommand{\eg}{\emph{e.g.,}\xspace}
\newcommand{\rf}{\emph{rf.}\xspace}

\newcommand{\ie}{\emph{i.e.,}\xspace}

\newcommand{\etal}{\emph{et al.}\xspace}

\AtBeginDocument{%
  \providecommand\BibTeX{{%
    \normalfont B\kern-0.5em{\scshape i\kern-0.25em b}\kern-0.8em\TeX}}}

\copyrightyear{2020}
\acmYear{2020}
\setcopyright{acmcopyright}
\acmConference[SIGIR '20]{Proceedings of the 43rd
International ACM SIGIR Conference on Research and Development in Information
Retrieval}{July 25--30, 2020}{Virtual Event, China}
\acmBooktitle{Proceedings of the 43rd International ACM SIGIR Conference on
Research and Development in Information Retrieval (SIGIR '20), July 25--30, 2020,
Virtual Event, China}
\acmPrice{15.00}
\acmDOI{10.1145/3397271.3401142}
\acmISBN{978-1-4503-8016-4/20/07}

\title{Global Context Enhanced Graph Neural Networks for Session-based Recommendation}
\author{Ziyang Wang\textsuperscript{1},
Wei Wei\textsuperscript{1,\XBox},
Gao Cong\textsuperscript{2},
Xiao-Li Li\textsuperscript{3},
Xian-Ling Mao\textsuperscript{4},
Minghui Qiu\textsuperscript{5}
}
\affiliation{\textsuperscript{1} Cognitive Computing and Intelligent Information Processing (CCIIP) Laboratory, School of Computer Science and Technology, Huazhong University of Science and Technology}
\affiliation{\textsuperscript{2} School of Computer Engineering, Nanyang Technological University, Singapore}
\affiliation{\textsuperscript{3} Institute for Infocomm Research, Singapore}
\affiliation{\textsuperscript{4} School of Computer Science and Technology, Beijing Institute of Technology}
\affiliation{\textsuperscript{5} Alibaba Group}
\affiliation{\textsuperscript{1} \{ziyang1997, weiw\}@hust.edu.cn\quad\textsuperscript{2} gaocong@ntu.edu.sg\quad \textsuperscript{3} xlli@i2r.a-star.edu.sg\quad\\ \textsuperscript{4} maoxl@bit.edu.cn \quad\textsuperscript{5} minghuiqiu@gmail.com}



\begin{document}
\begin{abstract}
\let\thefootnote\relax\footnotetext{\XBox: Corresponding Author}

Session-based recommendation (SBR) is a challenging task,
which aims at recommending 
items based on anonymous behavior sequences.
Almost all the existing solutions for SBR model user preference only based on the current session without exploiting the other sessions, which may contain both relevant and irrelevant item-transitions to the current session.
This paper proposes a novel approach, called \textbf{G}lobal \textbf{C}ontext \textbf{E}nhanced \textbf{G}raph Neural \textbf{N}etworks (\textsf{GCE-GNN}) to exploit item transitions over all sessions in a more subtle manner for better inferring the user preference of the current session.
Specifically, \textsf{GCE-GNN} learns two levels of item embeddings from \emph{session} graph and \emph{global} graph, respectively:
(i) \emph{Session graph}, which is to learn the session-level item embedding by modeling pairwise item-transitions within the current session;
and (ii) \emph{Global graph}, which is to learn the global-level item embedding by modeling pairwise item-transitions over all sessions.
In \textsf{GCE-GNN},
we propose a novel \emph{global-level item representation learning layer}, which employs a session-aware attention mechanism
to recursively incorporate the neighbors' embeddings of each node on the \emph{global} graph.
We also design a \emph{session-level item representation learning layer}, which employs a GNN on the \emph{session} graph to learn \emph{session}-level item embeddings within the current session.
Moreover, \textsf{GCE-GNN} aggregates the learnt item representations in the two levels
with a soft attention mechanism.
Experiments on three benchmark datasets demonstrate that GCE-GNN outperforms the state-of-the-art methods consistently.

\end{abstract}

\maketitle

{\fontsize{8pt}{8pt} \selectfont
\textbf{ACM Reference Format:} \\
Ziyang Wang, Wei Wei, Gao Cong, Xiao-Li Li, Xian-Ling Mao, Minghui Qiu. 2020. Global Context Enhanced Graph Neural Networks for Session-based Recommendation. In \textit{Proceedings of the 43rd International ACM SIGIR Conference on Research and Development in Information Retrieval (SIGIR ’20), July 25–30, 2020, Virtual Event, China.} ACM, New York, NY, USA, 10 pages. \url{https://doi.org/10.1145/3397271.3401142} }

\input{Section/Introduction}

\input{Section/RelatedWork}

\input{Section/Preliminaries}

\input{Section/TheProposedMethod}

\input{Section/Experiment}

\input{Section/Conclusion}

\section*{Acknowledgments}
This work was supported in part by the National Natural Science Foundation of China under Grant No.61602197 and Grant No.61772076, and in part by Equipment Pre-Research Fund for The 13th Five-year Plan under Grant No.41412050801.

\bibliographystyle{ACM-Reference-Format}
\balance
\bibliography{sample-sigconf}


\end{document}

%% file: Section/Introduction.tex
\section{Introduction}

Recommendation systems play critical roles on various on-line platforms, due to their success in addressing information overload problem by recommending useful content to users.
Conventional recommendation approaches (\eg collaborative filtering~\cite{sarwar2001item})
usually rely on the availability of user profiles and long-term historical interactions,
and may perform poorly in many recent 
real-world scenarios, \eg \emph{mobile stream media} like YouTube\footnote{\url{https://www.youtube.com/ }.} and Tiktok\footnote{\url{https://www.tiktok.com/}.},
when such information is unavailable (\eg unlogged-in user) or limited available (\eg short-term historical interaction).
Consequently, session-based recommendation has attracted extensive attention  recently,
which predicts the next interested item based on a given anonymous behavior sequence in chronological order.

Most of early studies on session-based recommendation  
fall into two categories, \ie
similarity-based~\cite{sarwar2001item} and chain-based~\cite{shani2005mdp}.
The former heavily replies on the co-occurrence information of items in the current session while neglecting
the sequential behavior patterns. The later infers all possible sequences of user choices over all items, which may suffer from intractable computation problem for real-world applications where the number of items is large. 
Recently, many deep learning based approaches are proposed for the task, which make use of pairwise item-transition information to model the user preference of a given session~\cite{hidasi2015session,li2017neural,ijcai2019-523,wang2019collaborative,kang2018self,ijcai2019-513}.
\begin{figure}[!t]
  \centering
  \includegraphics[width=\columnwidth, angle=0]{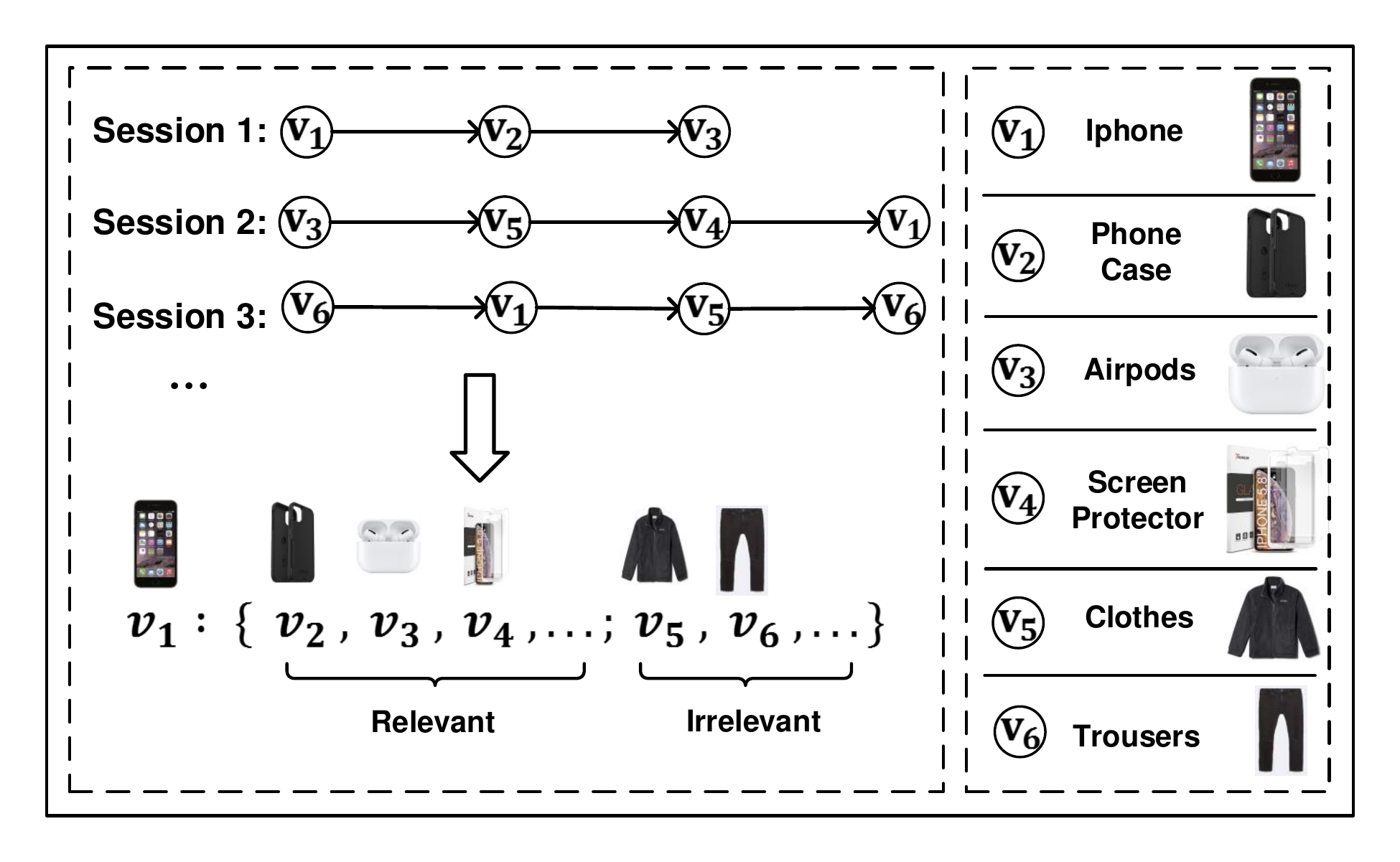}
  \caption{A toy example of \emph{global}-level item transition modeling.}
  \label{fig:frameWorkexample}
\end{figure}
These approaches have achieved encouraging results, but they still face the following issues.
{\emph{First}}, some of them infer the anonymous user's preference by sequentially extracting the session's pairwise item-transition information in chronological order using recurrent neural networks (RNN) (\eg GRU4REC~\cite{hidasi2015session}, NARM~\cite{li2017neural}) and memory networks~(\eg STAMP~\cite{liu2018stamp}).
However, a session may contain multiple user choices and even noise, and thus they may be insufficient in generating all correct dependencies, 
which suffer from the inability of modeling the complicated inherent order of item-transition patterns in embedding.
%
%
{\emph{Second}}, the others are based on graph neural networks~~\cite{wu2019session,ijcai2019-547,li2015gated} with self-attention mechanisms such as SR-GNN~\cite{wu2019session}. They learn the representation of the entire session by calculating the relative importance based on the session's pairwise
item-transition between each item and the last one,
and the performance  heavily
rely on the relevance of the last item to the user preference of the current session.


Furthermore, almost all the previous studies model user preference only based on the current session
while ignoring the useful item-transition patterns from other sessions.
To the best of our knowledge,  CSRM~\cite{wang2019collaborative} is the only work incorporating collaborative information from
the latest $m$ sessions to enrich the representation of the current session in end-to-end manner.
CSRM treats sessions as the minimum granularity and measure similarities
between the current and the latest $m$ sessions to extract collaborative information.
However, it may unfortunately encode both relevant and irrelevant information of the other sessions into
the current session embeddings, which may even deteriorate the performance~\cite{ijcai2019-523}. 
We illustrate this with an example in Figure~\ref{fig:frameWorkexample}.
Without loss of generality,
suppose the current session is ``Session 2'', and the session-based recommendation aims to recommend the relevant accessories related to ``Iphone ''.
From Figure~\ref{fig:frameWork}, we observe that:
(i) Utilizing the item-transition of the other sessions might help model
the user preference of the current session.
For example, we can find  \emph{relevant} pairwise item-transition information
for Session 2 from ``Session 1'' and ``Session 3'',
\eg a new pairwise item-transition ``[Iphone, Phone Case]'';
and (ii) Directly utilizing the item-transition information of the \emph{entire} other session may introduce noise
when part of the item-transition information encoded in such session is not relevant to the current session.
For instance,
CSRM~\cite{wang2019collaborative} may also consider to utilize ``Session 3'' to help modeling the user preference of ``Session 2'' if ``Session 3'' is one of the latest $m$ sessions,
and it will introduce the \emph{irrelevant} items (\ie ``clothes'' and ``trousers'')
when learning ``Session 2'''s embedding as it treats ``Session 3'' as a whole without distinguishing relevant item-transition from irrelevant item-transition, which is challenging.

To this end, we propose a novel approach to exploit the item-transitions over all sessions in a more subtle manner
for better inferring the user preference of the current session for session-based recommendation, which is named \textbf{G}lobal \textbf{C}ontext \textbf{E}nhanced \textbf{G}raph \textbf{N}eural \textbf{N}etworks (\textsf{GCE-GNN}).
%
%
In \textsf{GCE-GNN}, we propose to learn two  levels of item embeddings from \emph{session} graph and \emph{global} graph, respectively:
(i) \emph{Session graph}, which is to learn the session-level item embedding by modeling pairwise item-transitions within the current session;
and (ii) \emph{Global graph}, which is to learn the global-level item embeddings by modeling pairwise item-transitions over sessions~(including the current session).
%
%
In \textsf{GCE-GNN},
we propose a novel \emph{global-level item representation learning layer}, which employs a session-aware attention mechanism
to recursively incorporate the neighbors' embeddings of each node on the \emph{global} graph.
%
We also design a \emph{session-level item representation learning layer}, which employs a GNN on the \emph{session} graph to learn \emph{session}-level item embeddings within the current session.
Moreover, \textsf{GCE-GNN}  aggregates the learnt item representations in the two levels
with a soft attention mechanism.


The main contributions of this work are summarized as follows:

\begin{itemize}
\item To the best of our knowledge, this is the first work of exploiting \emph{global}-level item-transitions over all sessions
to learn \emph{global}-level contextual information for session-based recommendation.
%
\item We propose a unified model to improve the recommendation performance of the current session by effectively leveraging the pairwise item-transition information from two levels of graph models, \ie \emph{session} graph and \emph{global} graph.
\item We also propose a position-aware attention to incorporate the reversed position information in item embedding, which shows the superiority performance for session-based recommendation.
\item We conduct extensive experiments on three real-world datasets, which demonstrate that \textsf{GCE-GNN} outperforms nine baselines including state-of-the-art methods.
\end{itemize}

%% file: Section/RelatedWork.tex
\section{Related Work}

\paratitle{Markov Chain-based SBR}.
Several traditional methods can be employed for SBR although they are not originally designed for SBR. For example,  markov Chain-based methods
map the current session into a Markov chain, and then
infer a user's next action based on the previous one.
%
%
Rendle \etal \cite{rendle2010factorizing} propose FPMC
to capture both sequential patterns and long-term user preference
by a hybrid method based on the combination of matrix factorization and first-order Markov chain for recommendation. It can be adapted for SBR by ignoring the user latent representation as it is not available for anonymous SBR.
However, MC-based methods usually  focus on modeling sequential transition of two adjacent  items.
In contrast, our proposed model  converts the sequentially item-transitions into graph-structure data for capturing the inherent order of item-transition patterns for SBR.

\paratitle{Deep-learning based SBR}.
In recent years, neural network-based methods that are capable of  modeling sequential data have been utilized for SBR.
Hidasi \etal~\cite{hidasi2015session} propose the first work called GRU4REC
to apply the RNN networks for SBR, which adopts a multi-layer Gated Recurrent Unit (GRU) to model item interaction sequences.
Then, Tan \etal \cite{tan2016improved} extend the method \cite{hidasi2015session}   by introducing data augmentation.
Li \etal \cite{li2017neural} propose NARM that
incorporates attention mechanism into stack GRU encoder to capture the more representative item-transition information for SBR.
Liu~\etal~\cite{liu2018stamp} propose an attention-based short-term memory networks (named STAMP)
to captures the user's current interest without using RNN. Both NARM and STAMP emphasize the importance of the last click by using attention mechanism.
Inspired by $Transformer$ \cite{vaswani2017attention}, SASRec \cite{kang2018self} stacks multiple layers to capture the relevance between items.
ISLF \cite{ijcai2019-799} takes into account the user's interest shift, and employs  variational auto-encoder (VAE) and RNN
to capture the user's sequential behavior characteristics for SBR.
MCPRN \cite{ijcai2019-523} proposes to model the multi-purpose of a given session by using a mixture-channel model for SBR.
%
%
However, similar to MC-based methods, RNN-based methods focus on modeling the sequential transitions of adjacent items~\cite{ijcai2019-883}
to infer user preference via the chronology of the given sequence, and thus cannot model the complex item-transition patterns (e.g., non-adjacent item transitions).

Recently, several proposals  employ GNN-based model on graph built from the current session
to learn item embeddings for SBR.
Wu~\etal~\cite{wu2019session} propose a gated GNN model (named SR-GNN)
to learn item embeddings on the session graph, and then
obtain a representative session embedding by integrating each learnt item embedding with attentions,
which is calculated according to the relevance of each item to the last one.
Following the success of SR-GNN, some variants are also proposed for SBR,
such as GC-SAN \cite{ijcai2019-547}.
%
Qiu~\etal \cite{qiu2019rethinking} propose FGNN to learn each item representation by aggregating
its neighbors' embeddings with multi-head attention, and generate the final session representation
by repeatedly combining each learnt embeddings with the relevance of each time to the session.
However, all these approaches only model the item-transition information on the current session.
In contrast, our proposed model learns the item-transition information
over all sessions to enhance the learning from the current session.

\paratitle{Collaborative Filtering-based SBR}.
Although deep learning based methods have achieved remarkable performance,
collaborative filtering (CF) based methods can still provide competitive results.
Item-KNN \cite{sarwar2001item} can be extended for SBR by recommending items that are most similar to the last item of the current session.
%
KNN-RNN~\cite{jannach2017recurrent} makes use of GRU4REC~\cite{hidasi2015session}
and the co-occurrence-based KNN model to extract the sequential patterns for SBR.
Recently, Wang~\etal \cite{wang2019collaborative} propose an end-to-end neural model
named CSRM, which achieves state-of-the-art performance.
It first utilizes NARM over item-transitions to encode each session,
then enriches the representation of the current session by exploring the latest $m$ neighborhood sessions,
and finally utilizes a fusion gating mechanism to learn to combine different sources of features.
However,  it may suffer from noise when integrating other sessions' embeddings for the current one.
In contrast,  our proposed method considers the collaborative information in \emph{item-level}: we use the item embeddings in other sessions to enrich the item embeddings of the current session,
and then integrate them into session representation for SBR.

%% file: Section/Preliminaries.tex
\section{Preliminaries}
In this section, we first present the problem statement,
and then introduce two types of graph models, \ie \emph{session} graph
and \emph{global} graph,
based on different levels of pair-wise item transitions over sessions
for learning item representations,
in which we highlight the modeling of \emph{global}-level item transition
information as it is the basis of \emph{global} graph construction.

\subsection{Problem Statement}


Let  $V= \{ v_{1}, v_{2}, ..., v_{m}\}$ be all of items.
Each anonymous session, which is denoted by $S = \{ v^{s}_{1}, v^{s}_{2}, ..., v^{s}_{l}\}$, consists of a sequence of interactions (\ie items clicked by a user)
in chronological order, where $v^{s}_{i}$ denotes item $v_i$ clicked
within session $S$,
and the length of $S$ is $l$.

Given a session $S$, the problem of session-based recommendation aims to
recommend the top-$N$ items ($1\leq N \leq |V|$)
from $V$ that are most likely to be clicked by the user of the current session $S$.

\subsection{Graph Models: Session Graph and Global Graph}
In this subsection, we present two different  graph models to capture different levels of \emph{item transition } information
over all available sessions for item representation learning.

\subsubsection{Session Graph Model}
\begin{figure}[!t]
 \subfloat[\textbf{\emph{Session}} Graph.]{
  \begin{minipage}[b]{\linewidth}
  \centering
  \includegraphics[width=0.9\linewidth]{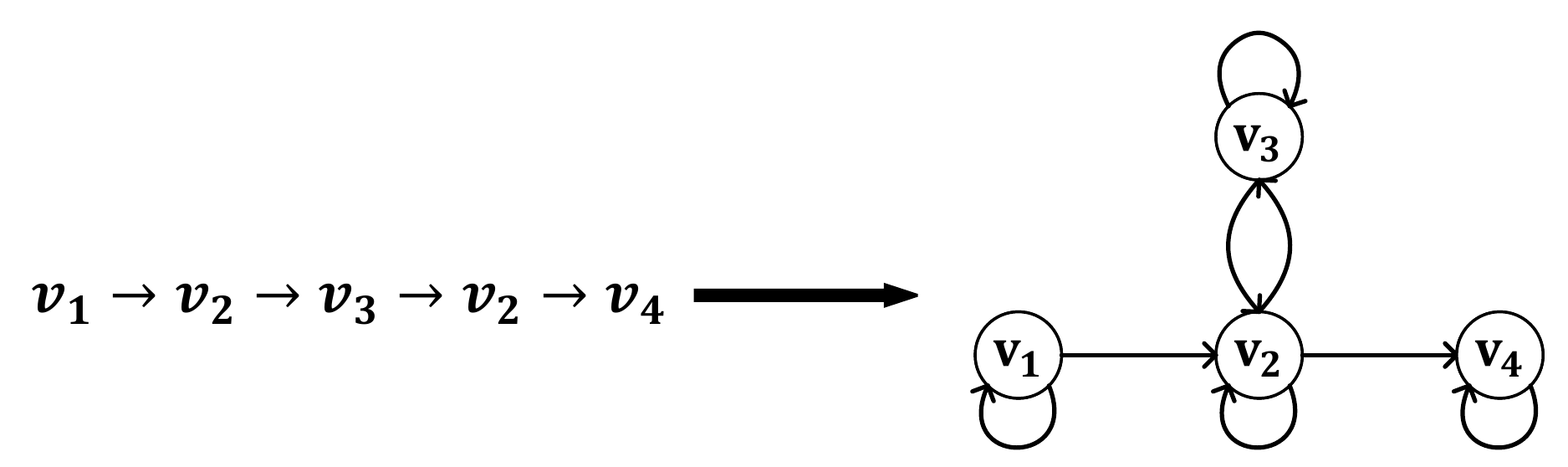}
  \end{minipage}
  \label{sessionGraph}
 }

 \subfloat[\textbf{\emph{Global}} Graph.]{
  \begin{minipage}[b]{\linewidth}
  \centering
  \includegraphics[width=0.9\linewidth]{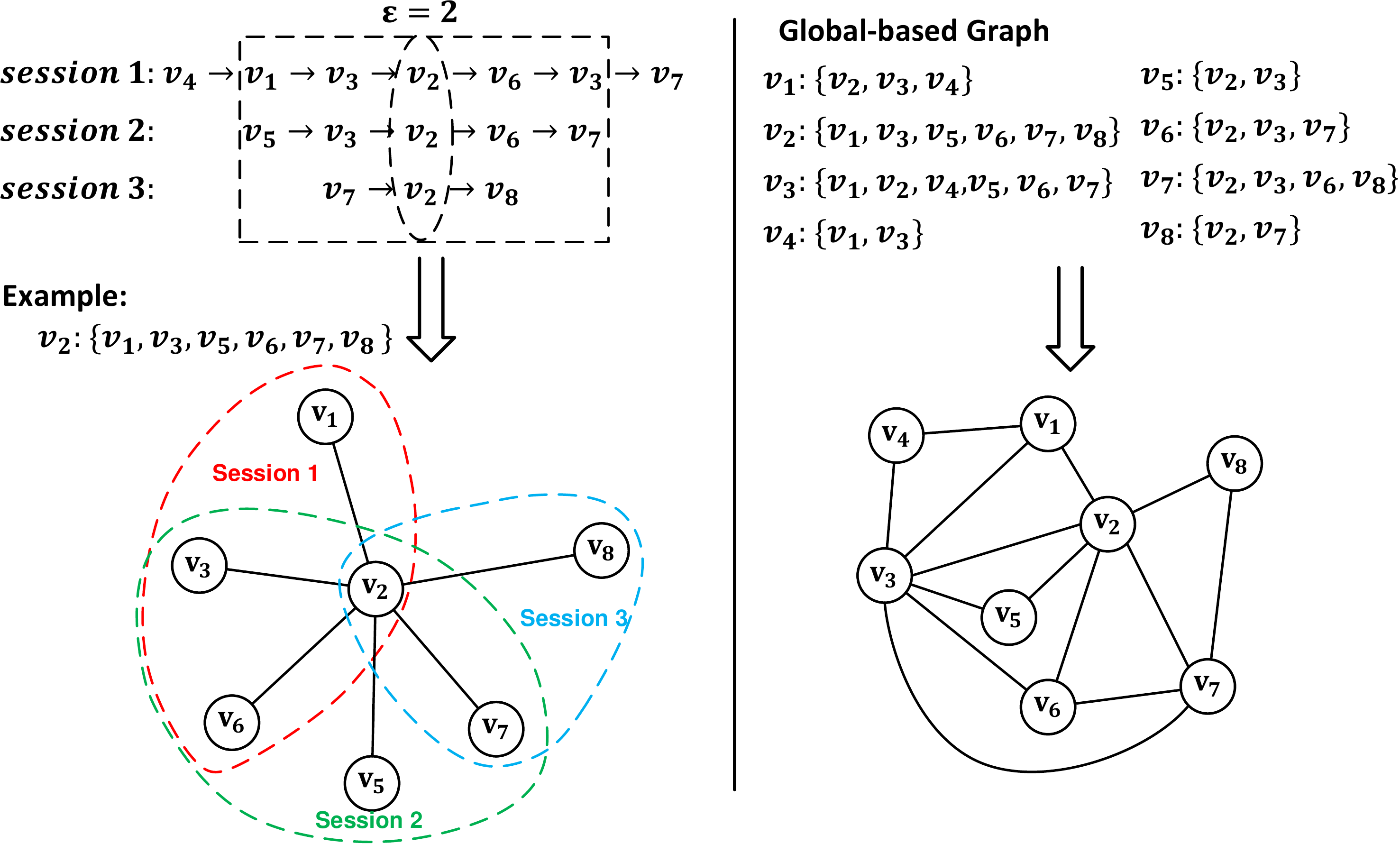}
  \end{minipage}
  \label{globalGraph}
 }
 \caption{Illustrations of construction of \emph{session} graph and  \emph{global} graph.}
\end{figure}

Session-based graph aims to learn the \textbf{\emph{session-level}} item embedding
by modeling sequential patterns over pair-wise adjacent items in the current session.
Inspired by \cite{wu2019session}, each session sequence
is converted into a session graph for learning the embeddings
of items in the current session via GNN,
which is defined as follows,
given session $S = \{ v^{s}_1, v^{s}_2, ..., v^{s}_l \}$,
let
\begin{math}
  \mathcal{G}_{s} = ( \mathcal{V}_{s}, \mathcal{E}_{s} )
\end{math}
be the corresponding \emph{session} graph,
where $\mathcal{V}_{s}\subseteq V$ is the set of clicked items in $S$,
$\mathcal{E}_{s}=\{e^{s}_{ij}\}$ denotes the edge set,
in which each edge indicates two adjacent items $(v^{s}_i, v^{s}_j)$ in $S$,
which is called \textbf{\emph{session}}-level item-transition pattern.
By following the work \cite{qiu2019rethinking},
each item is added a self loop (\rf Figure~\ref{sessionGraph}).

Different from \cite{wu2019session,qiu2019rethinking}, 
our session graph has four types of edges depending on the relationship between item $i$ and item $j$
which are denoted by $r_{in}$, $r_{out}$, $r_{in-out}$ and $r_{self}$. For edge $(v^s_i, v^s_j)$, $r_{in}$ indicates there is only transition from $v^s_j$ to $v^s_i$, $r_{out}$ implies there is only transition from $v^s_i$ to $v^s_j$, and $r_{in-out}$ reveals there are both transitions from $v^s_j$ to $v^s_i$ and from $v^s_i$ to $v^s_j$; $r_{self}$ refers to the self transition of an item.

%


%


\subsubsection{Global Graph Model}
Compared with traditional deep learning-based approaches (\eg RNN-based~\cite{li2017neural})
that focus on modeling sequential patterns of the entire session,
\emph{session} graph can efficiently capture complicated graph patterns of a session
to learn \emph{session}-level item embeddings.
%
%

However, we also aim to capture item-transition information from other sessions for learning representations of items, which is called \textbf{\emph{global}-level item transition information}.


\paratitle{Global-level Item Transition Modeling}.
Here,
we  take into account \emph{global}-level item transitions
for \emph{global}-level item representation learning, via integrating all pairwise item transitions over sessions.
%
As such, we propose a novel \emph{global} graph model for learning \emph{global}-level item embeddings,
which breaks down sequence independence assumption with linking all pairs of items
based on pairwise transitions over all sessions (including the current one).
Next, we firstly present a concept (\ie~{$\varepsilon$-neighbor set}) for modeling global-level item transition,
and then give the definition of \emph{global} graph.

\begin{definition}
\label{definition-1}
\textbf{$\varepsilon$-Neighbor Set}~($\mathcal{N}_{\varepsilon}({v})$).
\emph{For any item $v^{p}_i$ in session $S_p$, the $\varepsilon$-neighbor set of $v^{p}_i$ indicates a set of items,
each element of which is defined as follows,
{
\small
\begin{displaymath}
  \mathcal{N}_{\varepsilon}(v^{p}_i)=\left\{v^{q}_{j}| v^{p}_{i}=v^{q}_{i^{'}}\in S_p\cap S_q; v^{p}_{j}\in S_q; j\in [i^{'}-\varepsilon,i^{'}+\varepsilon]; S_p\neq S_q\right\},
\end{displaymath}
}
where $i^{'}$ is the order of item $v^{p}_i$ in session $S_q$,
$\varepsilon$ is a hyperparameter to control the scope of modeling of item-transition  between $v^{p}_i$
and the items in $S_q$.
Note that, parameter $\varepsilon$ favors the modeling of short-range item transitions
over sessions, since it is helpless (even noise, \eg irrelevant dependence) for
capturing the \emph{global}-level item transition information if beyond the scope ($\varepsilon$).
}
\end{definition}

According to Definition~\ref{definition-1}, for each item $v_i\in V$, \textbf{\emph{global}-level item transition}
is defined as
\begin{math}
  \{(v_i,v_j)|v_i,v_j\in V; v_j\in \mathcal{N}_{\varepsilon}({v_i})\}.
\end{math}
Notably, we do not distinguish the direction of \emph{global}-level item transition information for efficiency.

\paratitle{Global Graph}.
Global graph aims to capture the \emph{global}-level item transition information, which will be used
to learn item embeddings over all sessions.
%
%
Specifically, the  \emph{global} graph is built based on $\varepsilon$-neighbor sets
of items in all sessions.
Without loss of generality,
\emph{global} graph is defined as follows,
%
%
let $\mathcal{G}_{g} = ( \mathcal{V}_{g}, \mathcal{E}_{g} )$ be the \textbf{\emph{global}} graph,
where $\mathcal{V}_{g}$ denotes the graph node set that contains all items in $V$,
and
\begin{math}
\mathcal{E}_{g}=\{e^{g}_{ij}|(v_i,v_j)|v_i\!\in\! V, v_j\!\in\! \mathcal{N}_{\varepsilon}({v_i})\}
\end{math}
indicates the set of edges, each corresponding to two pairwise items from all the sessions.
Figure~\ref{globalGraph} shows an example of constructing the \emph{global} graph ($\varepsilon=2$).
Additionally, for each node $v_i$, we generate weight for its adjacent edges to distinguish the importance of $v_i$'s neighbors as follows: For each edge $(v_i$, $v_j)~(v_j\in \mathcal{N}^g_{v_i})$,
we use its frequency over all the sessions  as its weight of the corresponding edge; we only keep top-$N$ edges with the highest weights for each item $v_i$ on graph $\mathcal{G}_{g}$ due to  efficiency consideration.
%
%
Note that the definition of the neighbors\footnote{We do not distinguish $\mathcal{N}_{\varepsilon}({v})$
and $\mathcal{N}^{g}_{v}$ when the context is clear and discriminative.} (\ie $\mathcal{N}^{g}_{v}$) of item $v$ on graph $\mathcal{G}_{g}$
is same as $\mathcal{N}_{\varepsilon}(v)$.
Hence, $\mathcal{G}_{g}$ is  an \emph{undirected}  \emph{weighted} graph as $\varepsilon$-neighbor set is undirected.
During the testing phase, we do not \emph{dynamically} update the topological structure of \emph{global} graph
for efficiency consideration.
%

\paratitle{Remark}. Each item in $V$ is encoded into an unified embedding space at time-step $t$, \ie $\vec{h}^{t}_{i}\in\mathbb{R}^{d}$~($d$ indicates the dimension of item embedding),
which is feed with an initialization embedding $\vec{h}^{0}_{i}\in \mathbb{R}^{|V|}$,
here we use \emph{one-hot} based embedding
and it is transformed into $d$-dimensional latent vector space
by using a trainable matrix $\vec{W}_{0}\in \mathbb{R}^{d\times |V|}$.

%% file: Section/TheProposedMethod.tex
\section{The Proposed Method}

\begin{figure*}[t]
  \centering
  \includegraphics[width=1.6\columnwidth, angle=0]{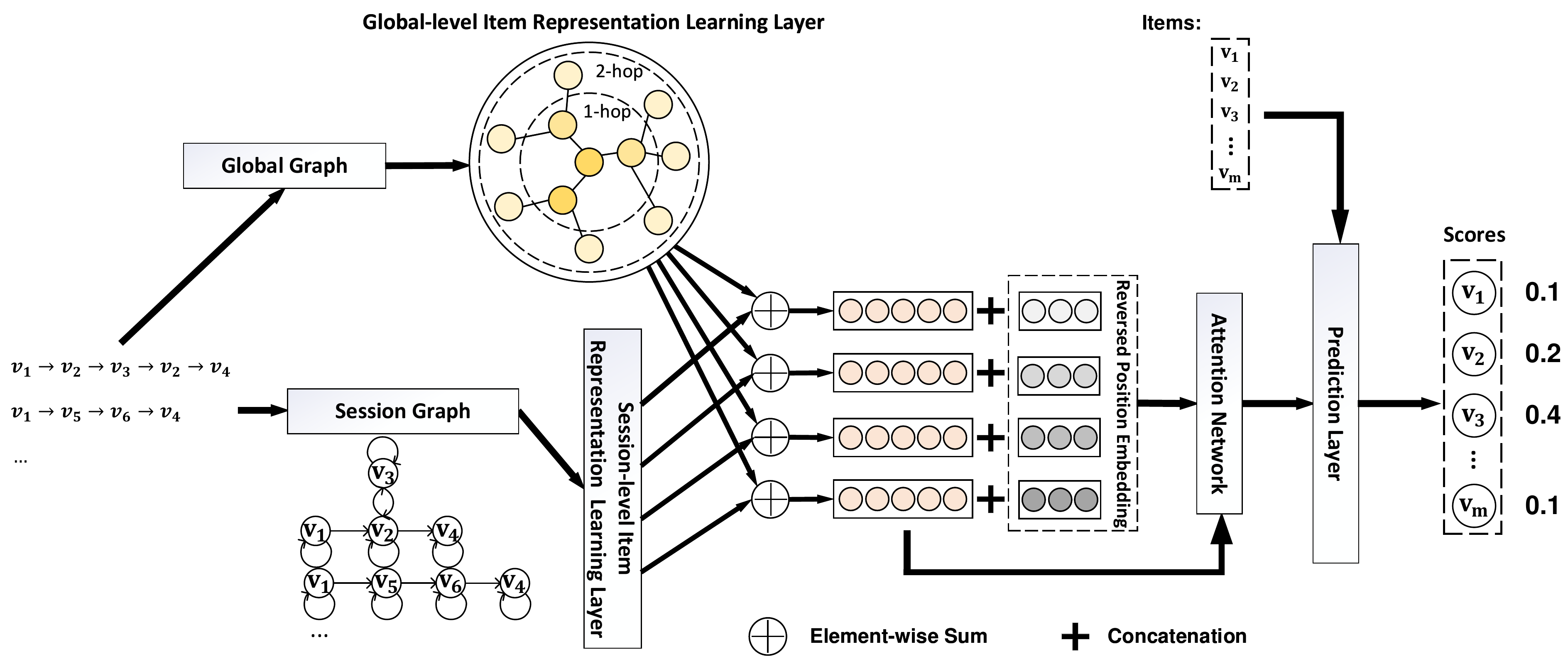}
  \vspace{-8pt}\caption{An overview of the proposed framework. Firstly, a global graph is constructed based on all training session sequences. Then for each session, a global feature encoder and local feature encoder will be used to extract node feature with global context and local context. Then the model incorporates position information to learn the contribution of each item to the next predicted item. Finally, candidate items will be scored.}
  \label{fig:frameWork}
\end{figure*}

We propose a novel \textbf{G}lobal \textbf{C}ontext \textbf{E}nhanced \textbf{G}raph \textbf{N}eural \textbf{N}etworks  for Session-based Recommendation (\textsf{GCE-GNN}).
%
 \textsf{GCE-GNN} aims to exploit both \emph{session}-level  and \emph{global}-level pairwise item transitions for  modeling the user preference of the current session for recommendation.
%
%
%
%
Figure~\ref{fig:frameWork} presents the architecture of \textsf{GCE-GNN}, which comprises four main components:
1) \emph{global-level item representation learning layer}. It learn \emph{global}-level
item embeddings over all sessions by employing a session-aware attention mechanism
to recursively incorporate each node's neighbors' embeddings
based on the global graph ($\mathcal{G}_{g}$) structure;
2) \emph{session-level item representation learning layer}. It employs a GNN model on session graph $\mathcal{G}_{s}$
to learn \emph{session}-level item embeddings within the current session;
3) \emph{session representation learning layer} It models the user preference of the current
session by aggregating the learnt item representations in both session-level and global-level;
4) \emph{prediction layer}. It outputs the predicted probability of candidate items
for recommendation.
We next present the four components in detail.

\subsection{Global-level Item Representation Learning Layer}


We next present how to propagate features on \emph{global} graph to encode item-transition information from other sessions to help recommendation.

Our layers are built based on the architecture of graph convolution network \cite{kipf2016semi}, and we generate the attention weights based on the importance of each connection by exploiting the idea of graph attention network \cite{velivckovic2017graph}. Here, we first describe a single layer, which consists of two components: \emph{information propagation} and \emph{information aggregation}, and then show how to generalize it to multiple layers.

\paratitle{Information Propagation}:
An item may be involved in multiple sessions, from which we can obtain useful item-transition information to effectively help current predictions.

To obtain the first-order neighbor's features of item $ v $,
one straightforward solution is to use mean pooling  method \cite{hamilton2017inductive}.
However, not all of items in $ v $'s $\varepsilon$-neighbor set are relevant to the user preference of the current session,
and thus we consider to utilize a session-aware attention to distinguish the importance of items in ($\mathcal{N}_{\varepsilon}(v)$).
Therefore, each item in $\mathcal{N}_{\varepsilon}(v)$ is linearly combined according to the session-aware attention score,
\begin{equation}
\vec{h}_{\mathcal{N}^g_{v_i}} = \sum_{v_j \in \mathcal{N}^g_{v_i}} \pi (v_i, v_j) \vec{h}_{v_j} ,
\end{equation}
where $\pi (v_i, v_j)$ estimates the importance weight of different neighbors. Intuitively, the closer an item is to the preference of the current session, the more important this item is to the recommendation. Therefore we implement $\pi (v_i, v_j)$ as follows:
\begin{equation}
\pi (v_i, v_j) = \vec{q}^T_1 \text{LeakyRelu} \big ( \matrix{W}_1 [ (\vec{s} \odot \vec{h}_{v_j}) \| w_{ij} ] \big ),
\end{equation}
here we choose \text{LeakyRelu} as activation function, $\odot$ indicates element-wise product, $\|$ indicates concatenation operation, $w_{ij} \in \mathbb{R}^{1}$ is the weight of edge $(v_i, v_j)$ in global graph, $\matrix{W}_1 \in \mathbb{R}^{d+1 \times d+1}$ and $\vec{q}_1 \in \mathbb{R}^{d+1}$ are trainable parameters, $\vec{s}$ can be seen as the features of current session, which is obtained by computing the average of item representations of the current session,
\begin{equation}
\vec{s} = \frac{1}{| S |} \sum_{v_i \in S} \vec{h}_{v_i}.
\label{eq:entire seesion information}
\end{equation}
Distinct from mean pooling, our approach makes the propagation of information dependent on the affinity between $ S $ and $ v_{j} $, which means neighbors that match the preference of current session will be more favourable.

Then we normalize the coefficients across all neighbors connected with $v_i$ by adopting the softmax function:
\begin{equation}
\pi (v_i, v_j) = \frac{\exp \big( \pi (v_i, v_j) \big)}{\sum_{v_k \in \mathcal{N}^g_{v_i}} \exp \big( \pi (v_i, v_k) \big)}.
\end{equation}
As a result, the final attention score is capable of suggesting which neighbor nodes should be given more attention.

\paratitle{Information Aggregation}: The final step is to aggregate the item representation $\vec{h}_{v}$ and its neighborhood representation $ h^g_{\mathcal{N}_v} $, we implement aggregator function \text{agg} as follows,
\begin{equation}
\vec{h}^{g}_{v} = \text{relu} \big( \matrix{W}_{2} [ \vec{h}_{v} \| \vec{h}_{\mathcal{N}^g_v} ]  \big),
\end{equation}
where we choose \text{relu} as the activation function and $\matrix{W}_{2} \in \mathbb{R}^{d \times 2d}$ is transformation weight.

Through a single aggregator layer, the representation of an item is dependent on itself and its immediate neighbors. We could explore the high-order connectivity information through extending aggregator from one layer to multiple layers, which allows more information related to the current session to be incorporated into the current representation. We formulate the representation of an item in the $k$-th steps as:
\begin{equation}
\vec{h}^{g, (k)}_{v} = \text{agg} \big ( \vec{h}^{(k-1)}_{v}, \vec{h}^{(k-1)}_{\mathcal{N}^g_{v}} \big ),
\end{equation}
$\vec{h}^{(k-1)}_{v}$ is representation of item $v$ which is generated from previous information propagation steps, $\vec{h}^{(0)}_{v}$ is set as $\vec{h}_{v}$ at the initial  propagation iteration. In this way, the $k$-order representation of an item is a mixture of its initial representations and its neighbors up to $k$ hops away. This enables more effective messages to be incorporated into the representation of the current session.


\subsection{Session-level Item Representation Learning layer}
The session graph contains pairwise item-transitions within the current session. We next present how to learn the session-level item embedding.

As the neighbors of item in session graph have different importance to itself, we utilize attention mechanism to learn the weight between different nodes. The attention coefficients can be computed through \emph{element-wise product} and \emph{non-linear transformation}:
\begin{equation}
e_{ij} = \text{LeakyReLU} \left( \vec{a}^{\top}_{r_{ij}} \left( \vec{h}_{v_i} \odot \vec{h}_{v_j }\right) \right),
\end{equation}
where $e_{ij}$ indicates the importance of node $v_j$'s features to node $v_i$ and we choose \text{LeakyReLU} as activation function, $r_{ij}$ is the relation between $v_i$ and $v_j$ and $\vec{a_{*}} \in \mathbb{R}^{d}$ are weight vectors.

For different relations, we train four weight vectors, namely $a_{in}$, $a_{out}$, $a_{in-out}$ and $a_{self}$. As not every two nodes are connected in the graph, we only compute $e_{ij}$ for nodes $j \in \mathcal{N}^{s}_{v_i}$ to inject the graph structure into the model, where $\mathcal{N}^{s}_{v_i}$ is the first-order neighbors of $v_i$. And to make coefficients comparable across different nodes, we normalize the attention weights through softmax function:
\begin{equation}
\label{attention coefficients}
\alpha_{i j}=\frac{\exp \left( \text{ LeakyReLU }\left(\vec{a}^{\top}_{r_{ij}} \left(\vec{h}_{v_i} \odot \vec{h}_{v_j} \right)\right)\right)}{\sum_{v_k \in \mathcal{N}^{s}_{v_i}} \exp \left( \text{LeakyReLU}\left(\vec{a}^{\top}_{r_{ik}} \left(  \vec{h}_{v_i} \odot \vec{h}_{v_k} \right)\right)\right)}.
\end{equation}
In Eq. (\ref{attention coefficients}) the attention coefficients $\alpha_{i j}$ is asymmetric, as their neighbors are different, which means the contribution they make to each other are unequal. Next we obtain the output features for each node by computing a linear combination of the features corresponding to the coefficients:
\begin{equation}
\vec{h}^{s}_{v_i} = \sum_{v_j \in \mathcal{N}^{s}_{v_i}} \alpha_{i j} \vec{h}_{v_j}.
\end{equation}
The item representations in session graph is aggregated by the features of item itself and its neighbor in the current session. Through the attention mechanism, the impact of noise on the session-level item representation learning is reduced.

\subsection{Session Representation Learning Layer}
For each item, we obtain its representations by incorporating both  global context  and session context, and its final representation is computed by sum pooling,
\begin{equation}
\begin{split}
\vec{h}^{g, (k)}_{v} &= \text{dropout} \big ( \vec{h}^{g, (k)}_{v} \big ) \\
\vec{h}_{v}^{\prime} &= \vec{h}^{g, (k)}_{v} + \vec{h}^{s}_{v},
\label{eq:combination}
\end{split}
\end{equation}
here we utilize \text{dropout}\cite{srivastava2014dropout} on global-level representation to avoid overfitting.

Based on the learnt item representations, we now present how to obtain the session representations. Different from previous work \cite{liu2018stamp,wu2019session,ijcai2019-547} which mainly focus on the last item, in this paper we propose a more comprehensive strategy to learn the contribution of each part of the session for prediction.

In our method, a session representation is constructed based on all the items involved in the session. Note that the contribution of different items to the next prediction is not equal.
Intuitively, the items clicked later in the session are more representative of the user's current preferences, which shows their greater importance for the recommendation. Moreover, it is important to find the main purpose of the user and filter noise in current session \cite{li2017neural}. Hence we incorporate reversed position information and session information to make a better prediction.

After feeding a session sequence into graph neural networks, we can obtain the representation of the items involved in the session, \ie $\matrix{H} = \left[ \vec{h}_{v^s_1}^{\prime}, \vec{h}_{v^s_2}^{\prime}, ..., \vec{h}_{v^s_l}^{\prime} \right]$. We also use a learnable position embedding matrix $\matrix{P} = \left[ \vec{p}_{1}, \vec{p}_{2}, ..., \vec{p}_{l} \right]$, where $\vec{p}_{i} \in \mathbb{R}^d$ is a position vector for specific position $i$ and $l$ is the length of the current session sequence. The position information is integrated through \emph{concatenation} and \emph{non-linear transformation}:
\begin{equation}
\vec{z}_{i} = \text{tanh} \left( \matrix{W}_{3}\left[ \vec{h}_{v^s_i}^{\prime} \parallel \vec{p}_{l-i+1} \right] + \vec{b}_{3} \right),
\end{equation}
where parameters $\matrix{W}_{3} \in \mathbb{R}^{d \times 2d}$ and $\vec{b}_{3} \in \mathbb{R}^{d}$ are trainable parameters.
Here we choose the reversed position embedding because the length of the session sequence is not fixed. Comparing to forward position information, the distance of the current item from the predicted item contains more effective information, \eg in the session $\{ v_2 \to v_3 \to ? \}$, $v_3$ is the second in the sequence and shows great influence to prediction, however in the session $\{ v_2 \to v_3 \to v_5 \to v_6 \to v_8 \to ? \}$, the importance of $v_3$ would be relatively small. Therefore the reversed position information can more accurately suggest the importance of each item.

The session information is obtained by computing the average of item representations of the session,
\begin{equation}
\vec{s}^{\prime} = \frac{1}{l} \sum_{i=1}^l \vec{h}_{v^s_i}^{\prime}.
\end{equation}
Next, we learn the corresponding weights through a soft-attention mechanism:
\begin{equation}
\beta_{i} = \vec{q}_2^{\top} \sigma \left( \matrix{W}_{4} \vec{z}_{i} + \matrix{W}_{5} \vec{s}^{\prime} + \vec{b}_{4} \right),
\end{equation}
where $\matrix{W}_{4}, \matrix{W}_{5} \in \mathbb{R}^{d \times d}$ and $\vec{q}_2, \vec{b}_{4} \in \mathbb{R}^{d}$ are learnable parameters.

Finally, the session representation can be obtained by linearly combining the item representations:
\begin{equation}
\vec{S} = \sum_{i=1}^l \beta_{i} \vec{h}_{v^s_i}^{\prime}.
\end{equation}

The session representation $\vec{S}$ is constructed by all the items involved in the current session, where the contribution of each item is determined not only by the information in the session graph, but also by the chronological order in the sequence.

\subsection{Prediction Layer}

Based on the obtained session representations $\vec{S}$, the final recommendation probability for each candidate item based on their initial embeddings as well as current session representation, and we first use dot product and then apply softmax function to obtain the output $\hat{y}$:
\begin{equation}
\hat{y}_i = \text{Softmax} \left( \vec{S}^{\top} \vec{h}_{v_i} \right),
\end{equation}
where $\hat{y}_{i} \in \hat{y}$ denotes the probability of item $v_{i}$ appearing as the next-click in the current session.

The loss function is defined as the cross-entropy of the prediction results $\hat{y}$:
\begin{equation}
\mathcal{L}(\hat{{y}}) = -\sum_{i=1}^{m} \vec{y}_{i} \log \left(\hat{y}_{i}\right)+\left(1-\vec{y}_{i}\right) \log \left(1-\hat{y}_{i}\right),
\end{equation}
where $\vec{y}$ denotes the one-hot encoding vector of the ground truth item.

%% file: Section/Experiment.tex
\section{Experiments}

We have conducted extensive experiments to evaluate the accuracy of the proposed GCE-GNN method by answering the following five key research questions:

\begin{itemize} [itemindent = 15pt]
\setlength{\itemsep}{3pt}

\item \noindent \textbf{RQ1}: Does GCE-GNN outperform  state-of-the-art SBR baselines in real world datasets?

\item \noindent \textbf{RQ2}: Does global graph and global-level encoder improve the performance of GCE-GNN? How well does GCE-GNN perform with different depth of receptive field $k$?

\item \noindent \textbf{RQ3}: Is reversed position embedding useful?

\item \noindent \textbf{RQ4}: How well does GCE-GNN perform with different aggregation operations?

\item \noindent \textbf{RQ5}: How do different hyper-parameter settings (\eg node dropout) affect the GCE-GNN's accuracy?

\end{itemize}

\subsection{Datesets and Preprocessing}
We employ three benchmark datasets, namely, $Diginetica$\footnote{https://competitions.codalab.org/competitions/11161}, $Tmall$\footnote{https://tianchi.aliyun.com/dataset/dataDetail?dataId=42} and $Nowplaying$\footnote{http://dbis-nowplaying.uibk.ac.at/\#nowplaying}. Particularly, Diginetica dataset is from CIKM Cup 2016, consisting of typical transaction data. 
Tmall dataset comes from IJCAI-15 competition, which contains anonymized user's shopping logs on Tmall online shopping platform. Nowplaying dataset comes from \cite{ismm14}, which describes the music listening behavior of users.

Following \cite{wu2019session,ijcai2019-547}, we conduct preprocessing step over the three datasets. 
More specifically, sessions of length 1 and items appearing less than 5 times were filtered across all the three datasets. Similar to \cite{liu2018stamp}, we set the sessions of last week (latest data) as the test data, and the remaining historical data for training. Furthermore, for a session $ S = \left[ s_{1}, s_{2}, ..., s_{n} \right]$, we generate sequences and corresponding labels by a sequence splitting preprocessing, i.e., ($\left[ s_1 \right], s_2$), ($\left[ s_1, s_2 \right], s_3$), ..., ($\left[ s_1, s_2,..., s_{n-1} \right], s_n$) for both training and testing across all the three datasets. The statistics of datasets, after preprocessing, are summarized in Table \ref{tab:statistic}.
{
\begin{table}[t]
	\centering
	\small
	\caption{Statistics of the used datasets.}
	\label{tab:statistic}
	\begin{tabular}{l|r|r|r}
	\hline
		{ \text{Dataset} } & \text{Diginetica} & \text{Tmall} & \text{Nowplaying}\\
		\hline
		{ \text{\# click} }  & 982,961 & 818,479 & 1,367,963 \\
		\hline
		{ \text{\# train} }  & 719,470 & 351,268 & 825,304 \\
		\hline
		{ \text{\# test} }  & 60,858 & 25,898 & 89,824 \\
		\hline
		{ \text{\# items} } & 43,097 & 40,728 & 60,417 \\
		\hline
		{ \text{avg. len.} } & 5.12 & 6.69 & 7.42 \\
	\hline
\end{tabular}
\end{table}
}

\subsection{Evaluation Metrics}
We adopt two widely used ranking based metrics: \textbf{P@N} and \textbf{MRR@N} by following previous work\cite{liu2018stamp, wu2019session}.

%
%


{
\setlength{\tabcolsep}{1pt}
\begin{table*}[t]
	\centering
	\caption{Effectiveness comparison on three datasets.}
    \small
	\label{tab:results}
	\begin{threeparttable}
	\begin{tabular}{c|cccc|cccc|cccc}
	\hline
		{ \textbf{Dataset} } & \multicolumn{4}{c|}{ \textbf{Diginetica} } & \multicolumn{4}{c|}{ \textbf{Tmall} } & \multicolumn{4}{c}{ \textbf{Nowplaying} } \\
		\hline
		{ \textbf{Methods} } & \textbf{P@10} & \textbf{P@20} & \textbf{MRR@10} & \textbf{MRR@20} & \textbf{P@10} & \textbf{P@20} & \textbf{MRR@10} & \textbf{MRR@20} & \textbf{P@10} & \textbf{P@20} & \textbf{MRR@10} & \textbf{MRR@20} \\
		\hline \hline
		{ \text{POP} } & 0.76 & 1.18 & 0.26 & 0.28 & 1.67 & 2.00 & 0.88 & 0.90 & 1.86 & 2.28 & 0.83 & 0.86 \\
		\hline
		{ \text{Item-KNN} } & 25.07 & 35.75 & 10.77 & 11.57 & 6.65 & 9.15 & 3.11 & 3.31 & 10.96 & 15.94 & 4.55 & 4.91 \\
		\hline
		{ \text{FPMC} } & 15.43 & 22.14 & 6.20 & 6.66 & 13.10 & 16.06 & 7.12 & 7.32 & 5.28 & 7.36 & 2.68 & 2.82 \\
		\hline \hline
		{ \text{GRU4Rec} } & 17.93 & 30.79 & 7.73 & 8.22 & 9.47 & 10.93 & 5.78 & 5.89 & 6.74 & 7.92 & 4.40 & 4.48 \\
		\hline
		{ \text{NARM} } & 35.44 & 48.32 & 15.13 & 16.00 & 19.17 & 23.30 & 10.42 &  10.70 & 13.6 & 18.59 & 6.62 & 6.93 \\
		\hline
		{ \text{STAMP} } & 33.98 & 46.62 & 14.26 & 15.13 & 22.63 & 26.47 & 13.12 & 13.36 & 13.22 & 17.66 & 6.57 & 6.88 \\
		\hline \hline
		{ \text{CSRM} } & 36.59 & 50.55 & 15.41 & 16.38 & \underline{24.54} & \underline{29.46} & \underline{13.62} & \underline{13.96} & 13.20 & 18.14 & 6.08 & 6.42 \\
		\hline \hline
		{ \text{SR-GNN} } & \underline{38.42} & 51.26 & \underline{16.89} & 17.78 & 23.41 & 27.57 & 13.45 & 13.72 & \underline{14.17} & \underline{18.87} & \underline{7.15} & \underline{7.47} \\
		\hline
		{ \text{FGNN} } & 37.72 & 50.58 & 15.95 & 16.84 & 20.67 & 25.24 & 10.07 & 10.39 & 13.89 & 18.78 & 6.8 & 7.15 \\
		{ \text{FGNN(reported)}\tnote{1}
		} & - & \underline{51.36} & - & \underline{18.47} & - & - & - & - & - & - & - & - \\
		\hline \hline
		{ \text{GCE-GNN} } & \textbf{41.16} & \textbf{54.22} & \textbf{18.15} & \textbf{19.04} & \textbf{28.01} & \textbf{33.42} & \textbf{15.08} & \textbf{15.42} & \textbf{16.94} & \textbf{22.37} & \textbf{8.03} & \textbf{8.40} \\
		{ \text{$p$-value} } & \textless0.001 & \textless0.001 & \textless0.001 & \textless0.001 & \textless0.001 & \textless0.001 & \textless0.001 & \textless0.001 & \textless0.001 & \textless0.001 & \textless0.001 & \textless0.01 \\
	\hline
\end{tabular}
\begin{tablenotes}
    \footnotesize
	\item[1] The codes of FGNN model released by the author are incomplete. For fairness, we compare our method with our re-implemented FGNN model as well as the results reported in original paper.
	
\end{tablenotes}
\end{threeparttable}
\end{table*}
}

\subsection{Baseline Algorithms}
We compare our method with classic methods as well as state-of-the-art models. The following nine baseline models are evaluated.

\paratitle{POP}: It recommends top-$N$ frequent items of the training set.

\paratitle{Item-KNN}\cite{sarwar2001item}: It recommends items based on the similarity between items of the current session and items of other ones.

\paratitle{FPMC}\cite{rendle2010factorizing}: It combines the matrix factorization and the first-order Markov chain for capturing both sequential effects and user preferences. By following the previous work, we also ignore the user latent representations when computing recommendation scores.

\paratitle{GRU4Rec}\footnote{https://github.com/hidasib/GRU4Rec} \cite{hidasi2015session}:
It is RNN-based model that uses Gated Recurrent Unit (GRU) to model user sequences.

\paratitle{NARM}\footnote{https://github.com/lijingsdu/sessionRec\_NARM} \cite{li2017neural}: It improves over \textbf{GRU4Rec}\cite{hidasi2015session} by incorporating attentions into RNN for SBR.

\paratitle{STAMP}\footnote{https://github.com/uestcnlp/STAMP} \cite{liu2018stamp}: It employs attention layers to replace all RNN encoders in previous work  by fully relying on the self-attention of the last item in the current session to capture the user's short-term interest.

\paratitle{SR-GNN}\footnote{https://github.com/CRIPAC-DIG/SR-GNN} \cite{wu2019session}: It employs a gated GNN layer to obtain item embeddings, followed by a self-attention of the last item as STAMP\cite{liu2018stamp} does to compute the session level embeddings for session-based recommendation.

\paratitle{CSRM}\footnote{https://github.com/wmeirui/CSRM\_SIGIR2019} \cite{wang2019collaborative}: It utilizes the memory networks to investigate the latest $m$ sessions for better predicting the intent of the current session.

\paratitle{FGNN}\footnote{https://github.com/RuihongQiu/FGNN} \cite{qiu2019rethinking}: It is recently proposed by designing a weighted attention graph layer to learn items embeddings, and the sessions for the next item recommendation are learnt by a graph level feature extractor.

\subsection{Parameter Setup}

Following previous methods \cite{li2017neural}\cite{liu2018stamp}\cite{wu2019session}, the dimension of the latent vectors is fixed  to $100$, and the size for mini-batch is set to $100$ for all models. We keep the hyper-parameters of each model consistent for a fair comparison. For CSRM, we set the memory size to 100 which is consistent with the batch size. For FGNN, we set the number of GNN layer to $3$ and the number of heads is set to $8$. For our model, all parameters are initialized using a Gaussian distribution with a mean of $0$ and a standard deviation of $0.1$. We use the Adam optimizer with the initial learning rate $0.001$, which will decay by $0.1$ after every $3$ epoch. The L2 penalty is set to $10^{-5}$ and the dropout ratio is searched in $\{ 0.1, 0.2, ..., 0.9 \}$ on a validation set which is a random $10\%$ subset of the training set. Moreover, we set the number of neighbors and the maximum distance of adjacent items $\varepsilon$ to $12$ and $3$, respectively.

\subsection{Overall Comparison (RQ1)}

Table \ref{tab:results} reports the experimental results of the 9 baselines and our proposed model on three real-world datasets, in which the best result of each column is highlighted in boldface. It can be observed that GCE-GNN achieves the best performance (statistically significant) across all three datasets in terms of the two metrics (with N=10, and 20) \textit{consistently}, which ascertains the effectiveness of our proposed method.

Among the traditional methods, POP's performance is the worst, as it only recommends top-$N$ frequent items. Comparing with POP, FPMC shows its effectiveness on three datasets, which utilizes first-order Markov chains and matrix factorization. Item-KNN achieves the best results among the traditional methods on the Diginetica and Nowplaying datasets. Note it only applies the similarity between items and does not consider the chronological order of the items in a session, and thus it cannot capture the sequential transitions between items.

Compared with traditional methods, neural network based methods usually have better performance for session-based recommendation. In sprite of preforming worse than Item-KNN on Diginetica, GRU4Rec, as the first RNN based method for SBR, still demonstrates the capability of RNN in modeling sequences.
However, RNN is designed for sequence modeling, and session based recommendation problems are not merely a sequence modeling task because the user’s preference may change within the
session.
%
%

The subsequent
methods, NARM and STAMP outperform GRU4REC
significantly. NARM combines RNN and attention mechanism, which uses the last hidden state of RNN as the main preference of user, this result indicates that directly using RNN to encode the session sequence may not be sufficient for SBR as RNN only models one way item-transition between adjacent items in a session. We also observe that STAMP, a complete attention-based method, achieves better performance than NARM on Tmall, which incorporates a self-attention over the last item of a session to model the short-term interest, this result demonstrates the effectiveness of assigning different attention weights on different items for session encoding. Compared with RNN, attention mechanism appears to be a better option, although STAMP neglects the chronological order of items in a session.

CSRM performs better than NARM and STAMP on Diginetica and Tmall. It shows the effectiveness of using item transitions from other sessions,
and also shows the shortcomings of the memory networks used by CSRM that have limited slots, additionally CSRM treats other sessions as a whole one
without distinguishing the relevant item-transitions from the irrelevant ones encoded in other sessions.

Among all the baseline methods, the GNN-based methods perform better on the Diginetica and Nowplaying datasets. By modeling every session sequence as a subgraph and applying GNN to encode items, SR-GNN and FGNN demonstrate the effectiveness of applying GNN in session-based recommendation.
This indicates that the graph modeling would be more
suitable than the sequence modeling, RNN, or a set
modeling, the attention modeling for SBR.

%

Our approach GCE-GNN outperforms SR-GNN and FGNN on all the three datasets. Specifically, GCE-GNN outperforms the SR-GNN by $ 6.86 \% $ on Diginetica, $16.34 \%$ on Tmall and $15.71 \%$ on Nowplaying on average.
Different from SR-GNN and FGNN, our approach integrates information from global context, i.e., other session, and local context, i.e., the current session, and also incorporates relative position information, leading to consistent better performance.


\subsection{Impact of Global Feature Encoder (RQ2)}
%
We next conduct experiments on three datasets to evaluate the effectiveness of global-level feature encoder and session-level feature encoder. Specially, we design four contrast models:
\begin{itemize}
\item GCE-GNN w/o global: GCE-GNN without global-level feature encoder and only with local feature
\item GCE-GNN w/o session: GCE-GNN without session-level feature encoder and only with global feature
\item GCE-GNN-$1$-hop: GCE-GNN with global-level feature encoder, which sets the number of hop to $1$.
\item GCE-GNN-$2$-hop: GCE-GNN with global-level  feature encoder, which sets the number of hop to $2$.
\end{itemize}
Table \ref{tab:resultGAT} shows the comparison between different contrast models. It is clear that with global-level feature encoder, GCE-GNN achieves better performance. Comparing with GCE-GNN w/o global context, GCE-GNN with $1$-hop and $2$-hop global-level feature encoder can explore item-transition information from other sessions, which helps the model to make more accurate predictions. It can also be observed that GCE-GNN with $2$-hop performs better than GCE-GNN with $1$-hop on Diginetica, indicating that high-level exploring might obtain more effective information from global graph. In addition, GCE-GNN with $1$-hop performs better than GCE-GNN with $2$-hop on Tmall, and this indicates that higher-level exploring might also introduce noise.

{
\setlength{\tabcolsep}{1pt}
\begin{table}[!t]
    \small
	\centering
	\caption{ The performance of contrast models.}
	\label{tab:resultGAT}
	\begin{tabular}{c|cc|cc|cc}
	\hline
		{ \text{Dataset} } & \multicolumn{2}{c|}{\text{Diginetica}} &  \multicolumn{2}{c|}{\text{Tmall}} &  \multicolumn{2}{c}{\text{Nowplaying}}\\
		\hline
		{ \text{Measures} }  & P@20 & MRR@20 & P@20 & MRR@20 & P@20 & MRR @20 \\
		\hline
		{ \text{w/o global} }  & 54.08 & 18.76 & 32.96 & 14.72 & \textbf{23.11} & 7.55 \\
		{ \text{w/o session} }  & 51.46 & 17.34 & 32.96 & 12.41 & 19.10 & 8.15 \\
		{ \text{1-hop} }  & 54.04 & 18.90 & \textbf{33.42} & \textbf{15.42} & 22.37 & \textbf{8.40} \\
		{ \text{2-hop} }  & \textbf{54.22} & \textbf{19.04} & 32.58 & 14.83 & 22.45 & 8.29 \\
	\hline
\end{tabular}
\end{table}
}

{
\setlength{\tabcolsep}{1pt}
\begin{table}[!t]
    \small
	\centering
	\caption{The performance of contrast models.}
	\label{tab:resultPosition}
	\begin{tabular}{c|cc|cc|cc}
	\hline
		{ \text{Dataset} } & \multicolumn{2}{c|}{\text{Diginetica}} &  \multicolumn{2}{c|}{\text{Tmall}} &  \multicolumn{2}{c}{\text{Nowplaying}}\\
		\hline
		{ \text{Measures} }  & P@20 & MRR@20 & P@20 & MRR@20 & P@20 & MRR @20 \\
		\hline
		{ \text{GCE-GNN-NP} }  & 50.45 & 17.65 & 31.16 & 14.71 & 19.42 & 6.05 \\
		{ \text{GCE-GNN-SA} }  & 51.68 & 17.94 & 25.80 & 12.94 & 21.40 & 7.18 \\
		{ \text{GCE-GNN} }  & \textbf{54.22} & \textbf{19.04} & \textbf{33.42} & \textbf{15.42} & \textbf{22.37} &  \textbf{8.40}\\
	\hline
\end{tabular}
\label{table: position}
\end{table}
}

{
\setlength{\tabcolsep}{1pt}
\begin{table}[!t]
    \small
	\centering
	\caption{Effects of different aggregation operations.}
	\label{tab:resultPosition}
	\begin{tabular}{c|cc|cc|cc}
	\hline
		{ \text{Dataset} } & \multicolumn{2}{c|}{\text{Diginetica}} &  \multicolumn{2}{c|}{\text{Tmall}} &  \multicolumn{2}{c}{\text{Nowplaying}}\\
		\hline
		{ \text{Measures} }  & P@20 & MRR@20 & P@20 & MRR@20 & P@20 & MRR @20 \\
		\hline
		{ \text{Gate Mechanism} }  & 53.84 & 18.83 & 32.80 & 15.33 & \textbf{22.47} & 7.83 \\
		{ \text{Max Pooling} }  & 47.69 & 16.44 & 31.87 & 15.39 & 19.13 & 6.71 \\
		{ \text{Concatenation} }  & 51.72 & 17.03 & 31.55 & 14.89 & 19.88 &  7.93\\
		{ \text{Sum Pooling} }  & \textbf{54.22} & \textbf{19.04} & \textbf{33.42} & \textbf{15.42} & 22.37 & \textbf{8.40} \\
	\hline
\end{tabular}
\label{table: aggregator}
\end{table}
}

\subsection{Impact of Position Vector (RQ3)}
The position vector is used to drive GCE-GNN to learn the contribution of each part in the current session. Although SASRec \cite{kang2018self} has injected forward position vector into the model to improve performance, we argue that forward position vector has very limited effect on the SBR task. To verify this and evaluate the effectiveness of using the position vector in a reverse order, which is proposed in GCE-GNN,  we design a series of contrast models:
\begin{itemize}
\item GCE-GNN-NP: GCE-GNN with forward position vector replacing the reverse order position vector.
\item GCE-GNN-SA: GCE-GNN with self attention function replacing the position-aware attention.
\end{itemize}
Table \ref{table: position} shows the performance of different contrast models. We observe that our attention network with reversed position embedding performs better than the other two variants.

GCE-GNN-NP does not perform well on all datasets. That is because the model cannot capture the distance from each item to the predicted item, which will mislead the model when training for sessions of various lengths.

GCE-GNN-SA performs better than GCE-GNN-NP on three datasets, indicating that the last item in a session contains the most relevant information for recommendation. However, it does not perform well on Tmall dataset, as it lacks a more comprehensive judgment of the contribution of each item.

Comparing with the two variants, reversed position embedding demonstrates its effectiveness. This confirms that the reversed position information can more accurately suggest the importance of each item. Moreover, though the attention mechanism, we filter the noise in the current session, which makes the model perform better.


\subsection{Impact of Aggregation Operations (RQ4)}
As we use local feature encoder and global feature encoder, it is meaningful to compare GCE-GNN with different aggregation operations, \ie , gating mechanism, max pooling and concatenation mechanism.

For gating mechanism, we use a linear interpolation between local feature representation $h^l$ and global feature representation $h^g$:
\begin{equation}
\begin{split}
\vec{r}_v &= \sigma ( \matrix{W}_s \vec{h}^{s}_{v} + \matrix{W}_g \vec{h}^{g}_{v}) \\
 \vec{h}_{v}^{\prime} &= \vec{r}_v \vec{h}^{g}_{v} + (\vec{1} - \vec{r}_v) \vec{h}^{s}_{v}, \\
\end{split}
\end{equation}
where $\sigma$ is the \text{sigmoid} activation function and $\vec{r}_v$ is learned to balance the importance of two features.

For max pooling, we take the maximum value of every dimension for each feature, and the $i-$th dimension of an item representation $\vec{h}_{vi}^{\prime}$ is formulated as
\begin{equation}
\vec{h}_{vi}^{\prime} = \text{max} (\vec{h}^{g}_{vi}, \vec{h}^{s}_{vi}).
\end{equation}

For the concatenation operation, the final representation is the concatenation of vectors $\vec{h}^{g}_{v}$ and $\vec{h}^{s}_{v}$:
\begin{equation}
\vec{h}_{v}^{\prime} = \matrix{M} \big( [ \vec{h}^{g}_{v} \| \vec{h}^{s}_{v} ] \big)
\end{equation}
where $\matrix{M} \in \mathbb{R}^{d \times 2d}$ is the transformer weight.

Table \ref{table: aggregator} shows the performance of different aggregation operations on the three datasets. It can be observed that GCE-GNN with sum pooling outperforms other aggregation operations on Diginetica and Tmall in terms of Recall@20 and MRR@20. Max pooling's performance is the worst on Diginetica but it performs better than the other two aggregators on Tmall in terms of MRR@20. Despite of using additional parameters, Gate mechanism and Concatenation's performance is also worse than sum pooling, possibly because too many parameters may lead to overfitting.

\begin{figure}[!t]
\begin{center}
\subfloat[{Diginetica}]{\includegraphics[width=0.2\textwidth,angle=0]{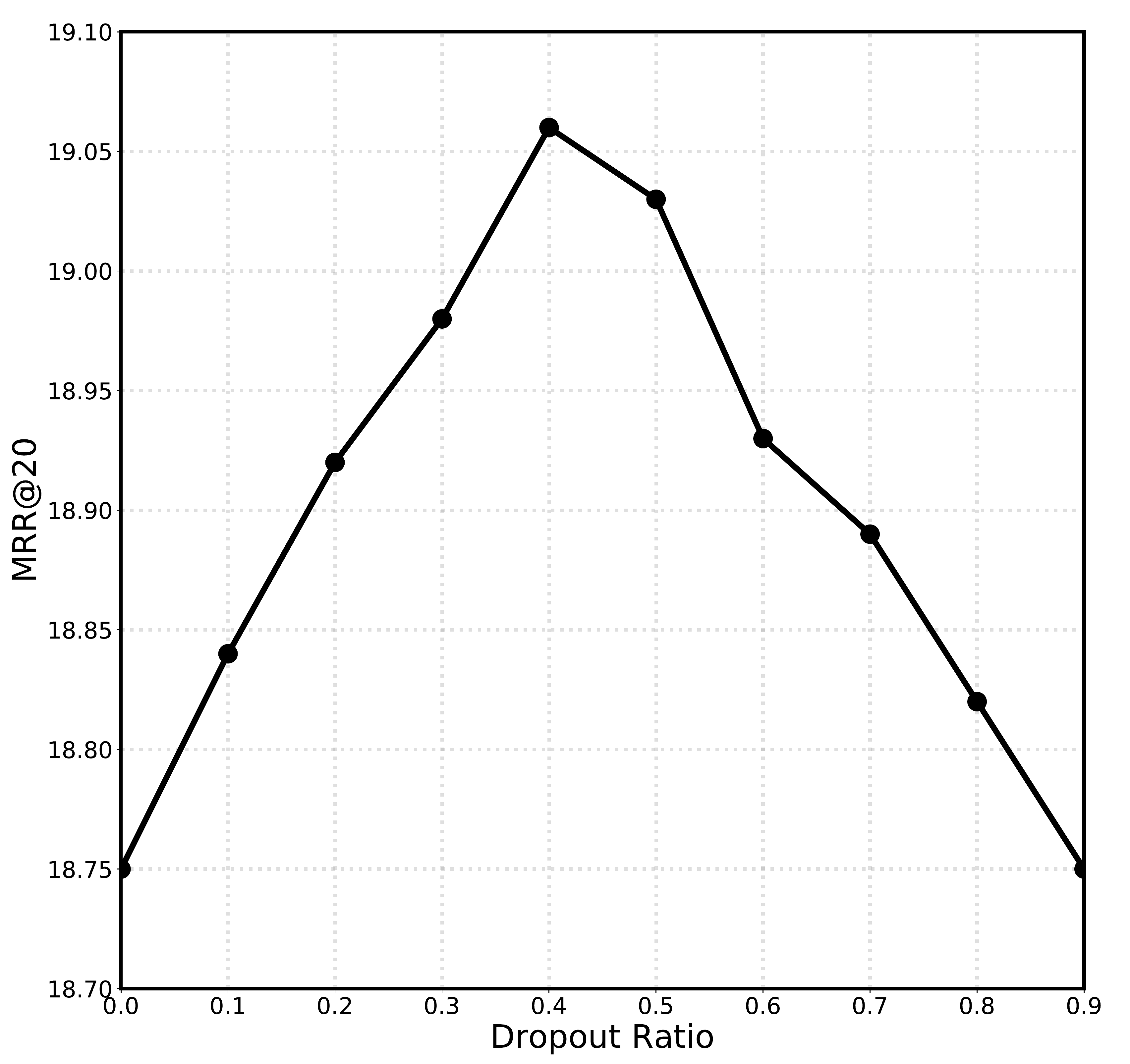}}\quad
\subfloat[{Tmall}]{\includegraphics[width=0.2\textwidth, angle=0]{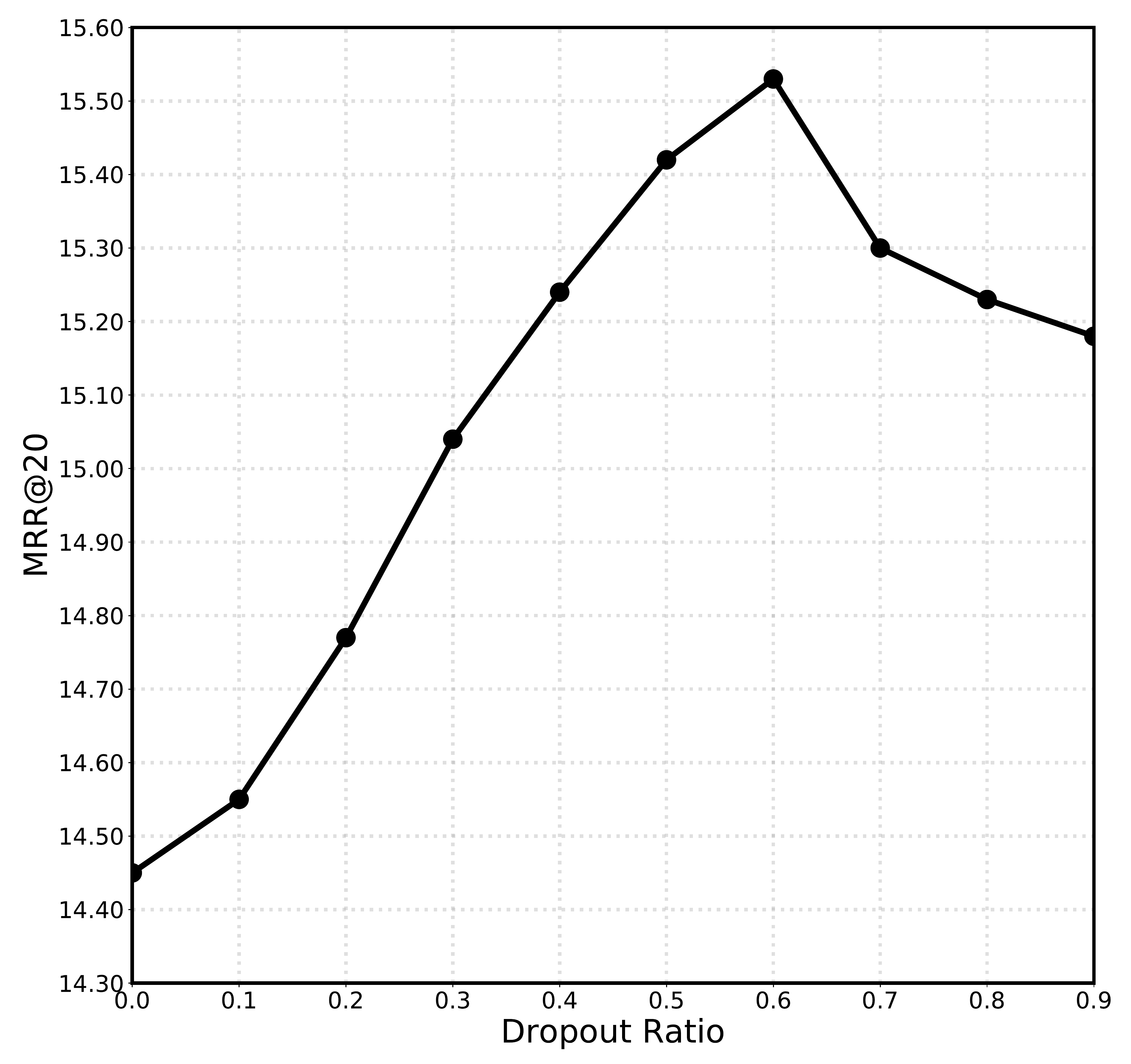}}
\caption{Impact of dropout ratio.}
\label{dropout}
\end{center}
\end{figure}

\subsection{Impact of Dropout Setting (RQ5)}
To prevent \text{GCE-GNN} from overfitting, we employ dropout \cite{srivastava2014dropout} regularization techniques, which have been shown to be effective in various neural network architectures including graph neural networks\cite{wang2019neural}\cite{wang2019heterogeneous}. The key idea of dropout is to randomly drop neurons with probability $p$ during training, while using all neurons for testing. Figure \ref{dropout} shows the impact of dropout in Equation (\ref{eq:combination}) on Diginetica and Tmall datasets. We can observe that when dropout ratio is small, the model does not perform well on both datasets, as it is easy to overfit. It achieves the best performance when dropout ratio is set to 0.4 on Diginetica and 0.6 on Tmall. However, when dropout ratio is big, the performance of the model starts to deteriorate, as it is hard for the model to learn from data with limited available neurons.

%% file: Section/Conclusion.tex
\section{Conclusion}
This paper studies the problem of session-based recommendation, which is a challenging task as 
the user identities and historical interactions are often unavailable due to privacy and data protection concern.
It proposes a novel architecture for session-based recommendation based on graph neural network. Specifically, it first converts the session sequences into session graphs and construct a global graph. The local context information and global context information are subsequently combined to enhance the feature presentations of items. Finally, it incorporates the reversed position vector and session information to empower the proposed model to better learn the contribution of each item. Comprehensive experiments demonstrate that the proposed method significantly outperforms nine baselines over three benchmark datasets consistently.